\documentclass[aps,prb,twocolumn,amsmath,amssymb,gallery,footinbib,superscriptaddress,floatfix]{revtex4-2}
\usepackage{graphicx}
\usepackage{dcolumn}
\usepackage{bm}
\usepackage{hyperref}
\usepackage{stackrel,amssymb}
\usepackage{placeins}
\hypersetup{colorlinks=true, linkcolor=blue, citecolor=blue, urlcolor=magenta, pdftitle={On the search of displacive chiral phase transitions}, pdfauthor={F. Gomez-Ortiz}}
\usepackage{bbding}
\usepackage{xcolor} 
\usepackage[normalem]{ulem}

\begin{document}
\preprint{APS/123-QED}
\title{Emergence of Transverse Dielectric Response in Ferroelectric Dielectric Heterostructures}
\author{Fernando G\'omez-Ortiz}
\email[]{fgomez@uliege.be}
\affiliation{Physique Théorique des Matériaux, QMAT, CESAM, Université de Liège, B-4000 Sart-Tilman, Belgium} 
\author{Ramamoorthy Ramesh}
\affiliation{Department of Materials Science and Engineering, University of California, Berkeley, California 94720,
USA}
\affiliation{Department of Materials Science and Nanoengineering, Rice University, Houston, Texas 77251,
USA}
\author{Javier Junquera}
\affiliation{Departamento de Ciencias de la Tierra y Física de la Materia Condensada, Universidad de Cantabria, Avenida de los Castros s/n 39005 Santander. Spain.}

\date{\today}
\begin{abstract}
We report the emergence of a transverse dielectric response in PbTiO$_{3}$/SrTiO$_{3}$ superlattices hosting polar vortex structures. Using second-principles simulations, we find that an electric field applied along one direction induces significant local polarization responses along orthogonal directions, with magnitudes approaching half that of the diagonal susceptibility components. These off-diagonal responses are strongly dependent on the topology of the vortex structure and can be deterministically tuned or even reversed via homogeneous electric fields or epitaxial strain. Notably, the transverse susceptibilities become comparable to the diagonal components during a field- or strain-induced transition to a polarization wave state. This discovery opens avenues for engineering reconfigurable nanoscale dielectric responses in topologically textured ferroelectric systems.
\end{abstract}
\maketitle
Complex polarization textures have recently emerged as a rich and exciting playground in condensed matter physics. Novel phases, arising from the delicate interplay between electrostatic, elastic, and gradient energies, have been uncovered in a variety of systems with particular prominence in ferroelectric/dielectric heterostructures~\cite{Junquera-23}. 
These include intriguing topologically nontrivial structures such as vortices~\cite{Yadav-16}, skyrmions~\cite{Das-19}, merons~\cite{Wang-20}, and hopfions~\cite{Luk-20}. 
Beyond their fundamental interest, these textures exhibit fascinating functional properties, such as enhanced electromechanical responses~\cite{Bellaiche-00}, chirality~\cite{Louis-12,Shafer-18} or ultra-fast and complex dynamic behavior~\cite{Daranciang-12,Li-21}.
Notably, one property that has garnered significant attention is the enhancement of their dielectric permittivity due to the emergence of local regions located at the vortex core (or skyrmion periphery)  showing negative permittivity~\cite{Bratkovsky-01,Luk-19NC,Zubko-16,Iniguez-19}. This phenomenon leads to voltage amplification~\cite{Graf-22}, offering a path toward more energy-efficient devices, which could have promising applications in low-power computing~\cite{Salahuddin-2008}.
This behavior was first experimentally observed in polar vortices~\cite{Yadav-19} precisely located at the vortex core where the system is significantly more responsive as it presents a higher local energy and a higher ionic susceptibility.

Remarkably, polar vortices in ferroelectric/dielectric heterostructures have also been the subject of other significant findings, including the discovery of collective dynamics with orders-of-magnitude-higher frequencies and smaller lateral sizes than those of experimentally realized magnetic vortices~\cite{Li-21}.
Their dynamics is marked by the presence of a buckling between the vortex cores and their up/down displacement towards the interface.
This movement is coupled with the onset of an in-plane polarization component and can therefore be stabilized by a slight tensile strain and has been shown to be controllable by homogeneous electric fields, offering pathways to manipulate the chiral nature of the system~\cite{Behera-22} and enabling the emergence of novel metastable switching behaviors, which can be exploited for the development of non-volatile memory devices~\cite{Behera-23}.

All these emergent phenomena are likely connected to the dielectric response of such topological features, particularly the nature of the dielectric tensor matrix elements in the presence of such polar textures. The formation of such polar textures (vortices and skyrmions) also reduces the overall crystal symmetry of the heterostructure from a cubic/tetragonal symmetry for the PbTiO$_{3}$/SrTiO$_{3}$ layers to an overall symmetry of ``$m$'' for the vortices and skyrmions. For example, this would be reflected in non-trivial values of the off-diagonal elements of the dielectric susceptibility of the individual layers and the heterostructure. In this article, we investigate the electric susceptibility response of the polar vortex phase in PbTiO$_3$/SrTiO$_3$ superlattices using second-principles calculations implemented in the {\sc{scale-up}} package~\cite{Wojdel-13}. We report the emergence of novel transverse responses, where the application of an electric field along one direction induces local polarization changes and the corresponding dielectric susceptibilities along perpendicular directions.  Furthermore, this response is highly tunable by external electric fields or tensile strain, allowing both the magnitude and even the sign to be adjusted at will. This phenomenon could have significant technological implications, following the revolution sparked by the negative capacitance effect, by providing new avenues for designing tunable devices with distinct dielectric responses.

\begin{figure}[tbhp]
     \centering
      \includegraphics[width=8.5cm]{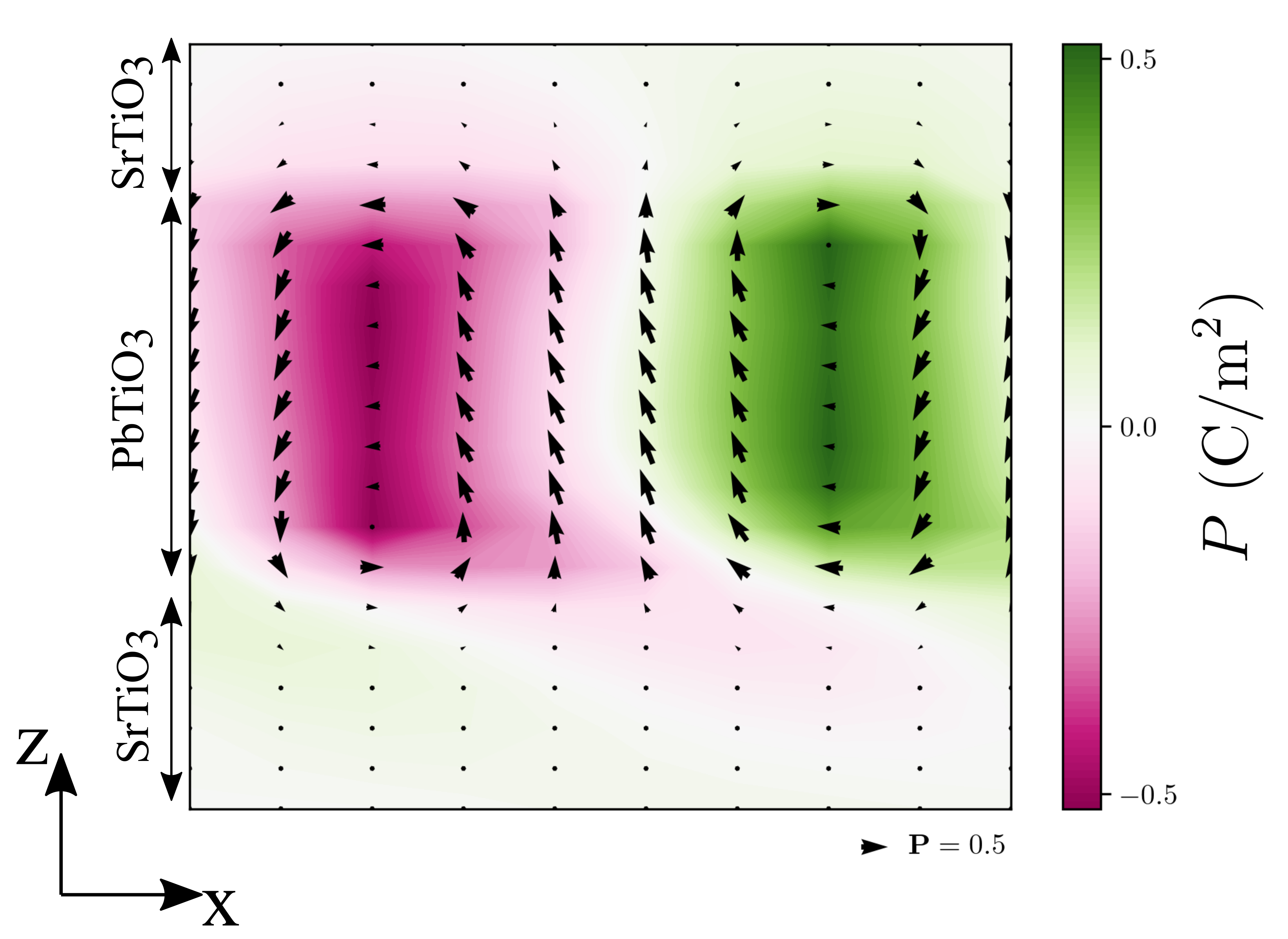}
      \caption{Planar view of the local dipole pattern in a cross-section of the infinite stripes. Arrows indicate the projection of the dipoles onto the ($x$, $z$) plane (bar at the bottom represents the scale), whereas colors refer to the perpendicular dipole component along $y$.}
      \label{fig:Fig1} 
\end{figure}

To this end, we consider a  superlattice with 10 unit cells of PbTiO$_3$ and 10 unit cells of SrTiO$_3$ under a slight tensile strain of $0.25\%$~\cite{Behera-22} with respect to the reference structure of ($a=b=3.901$~\AA) used in previous works~\cite{Zubko-16} as a representative model system (see End Matter of the letter for computational details).
Under such conditions, and in agreement with previous works~\cite{Behera-22}, we find that at low temperatures the system forms an ordered array of counter rotating vortices, where the polarization domains are oriented upwards and downwards showing a characteristic periodicity marked by the Kittel law~\cite{Kittel-49,kittel1946,Gomez-23} (see Fig.~\ref{fig:Fig1}). At the domain wall, the polarization profiles are closed by the vortex textures.
Due to the slight tensile strain on the structure, an in-plane component of the polarization and the concomitant buckling of the vortices is observed. In this state, the clockwise vortex can be located either at the bottom or the top of the structure, with both configurations being energetically degenerate and equally probable.
When the clockwise vortex is positioned at the bottom, the polarization along the $x$-direction is positive, while it becomes negative when the clockwise vortex is at the top.
Moreover, an axial component of the polarization along the $y$ direction can be observed. The existence of such a polarization component is explained by the ferroelectric nature of domain walls~\cite{Wojdel-14}, and although it is not ubiquitous~\cite{Zatterin-24}, its presence is enhanced by the elastic constraints to which the system is subjected~\cite{Chege-25}. Parallel or antiparallel configurations are numerically degenerate~\cite{Shafer-18}, however, to comply with most of the experimental evidence~\cite{Yadav-19,Damodaran-17.2} an antiparallel alignment has been selected as the initial configuration.
For computational feasibility, a superlattice with 2 unit cells along the $y$ directions has been chosen. This size is sufficiently large to allow octahedral rotations in the SrTiO$_3$ layers, which can affect the dielectric response of the material. Additionally, limiting the structure to only 2 layers along the $y$ direction helps speed up the calculations while maintaining the essential physics of the system as further explained in the End Matter of the letter.

To investigate the material's response to applied electric fields, we performed Monte Carlo simulations. By analyzing the local variations in polarization induced by small finite differences in the electric field ($\pm0.026$ MV/cm), we obtained spatially resolved measurements of the local electric susceptibility. These results reveal the intricate dielectric behavior of the system at the nanoscale, with distinct variations depending on the direction of the applied field and the corresponding polarization response.

\begin{figure*}[tbhp]
     \centering
      \includegraphics[width=17cm]{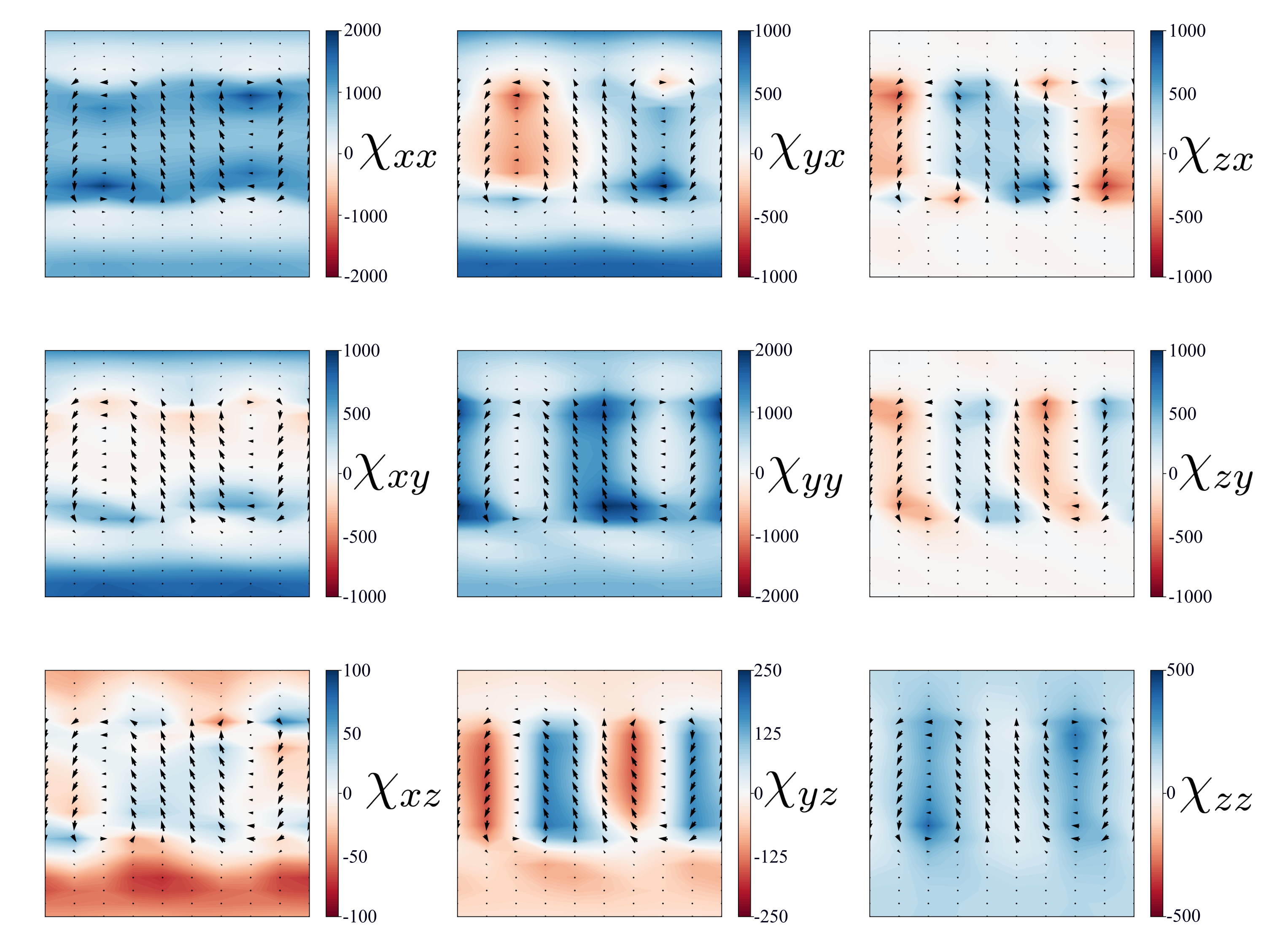}
      \caption{Local distribution of electric susceptibility responses obtained through finite differences in Monte Carlo simulations as explained in the End Matter of the letter. In each row, the system is subjected to electric fields applied along the $x$ (first row), $y$ (second row), and $z$ (third row) directions. The columns represent the resulting local dipole variations along the $x$ (first column), $y$ (second column), and $z$ (third column) directions. Arrows depict the projection of the dipoles onto the ($x$, $z$) plane from the original structure shown in Fig.~\ref{fig:Fig1}, while the color map illustrates the magnitude of the local susceptibility.}
      \label{fig:Fig2} 
\end{figure*}

A detailed visualization of these findings is presented in Fig.~\ref{fig:Fig2} where in each row, the system is subjected to electric fields applied along the $x$ (first row), $y$ (second row), and $z$ (third row) directions and the columns represent the resulting polarization variations along the $x$ (first column), $y$ (second column), and $z$ (third column) directions respectively. 
As expected, the diagonal elements, representing variations of the local polarization aligned with the electric field ($\partial P_i/\partial E_i$), are uniformly positive. This behavior arises because the dipoles align with the electric field to minimize the interaction energy $-\mathbf{E} \cdot \mathbf{P}$. In agreement with previous works~\cite{Yadav-19,Das-21}, this response is enhanced at domain wall regions~\cite{Zubko-16,Das-21}.

Remarkably, the off-diagonal matrix elements are of particular interest, as they exhibit a well-defined and structured pattern with susceptibility values that are substantial, typically around half of the diagonal terms. This stands in contrast to the random behavior with low values typically expected from fluctuations. This clear organization points to the emergence of local transverse dielectric responses, where the polarization not only aligns with the electric field but also responds in perpendicular directions. 

Let us first focus on the first row of Fig.~\ref{fig:Fig2} that shows the behavior of the $z$ and $y$ components of the polarization when an electric field along $x$ is applied on the sample.
As depicted in Fig.~\ref{fig:Fig1}, our original structure exhibits a negative polarization along the $x$ direction, arising from the offset between the vortex cores. When an electric field is applied along the $x$ direction, it leads to a reduction in this polarization component and a tendency to lessen the buckling of the vortex textures, thereby promoting the formation of more vertically aligned polarization domains. %
This behavior is reflected in the local susceptibility response observed in our system, where domains with positive (negative) polarization along $z$ show a positive (negative) susceptibility $\chi_{zx}$. Similarly, the axial polarization component, primarily located at the vortex core, also displays a slight negative polarization along $x$, as a consequence of the offset, that only becomes positive close to the bottom (top) interface to close the counter-clockwise (clockwise) vortex (see Fig.~\ref{fig:Fig1}). As the electric field is applied along $x$, this component decreases, giving way to an increase in the axial component's magnitude. This leads to the observed positive correlation between the positive and negative susceptibilities $\chi_{yx}$ and the positive and negative values of the axial polarization component in the original structure. 

A similar analysis can be conducted for the responses of $P_x$ and $P_y$ as a function of an electric field applied along the $z$-direction (see third row of Fig.~\ref{fig:Fig2}). 
Interestingly, due to the different nature of $x$- and axial-components and their role on the vortex texture, the associated dielectric tensors are not symmetric.
When an electric field is applied along $z$, the magnitude of the polarization in positively polarized domains increases, while it decreases in negatively polarized domains. 
Consequently, positively polarized regions experience a reduction in the magnitude of polarization along the $x$ and $y$ directions, whereas the opposite occurs in negatively polarized domains. 
This is evident from the blue regions of  $\chi_{xz}$ in the positively polarized domains, where the dipoles along the $x$-direction become smaller and hence less negative, and the red regions in the negatively polarized domains, where the dipoles along the $x$-direction become larger and hence more negative.
In short, when an electric field induces a local decrease (increase) in the polarization magnitude along the direction of the field, the dipole magnitude along the orthogonal directions increases (decreases) respectively. This behavior can also be observed in the $\chi_{xy}$ and $\chi_{zy}$ subplots.
\begin{figure}[btp]
     \centering
      \includegraphics[width=\columnwidth]{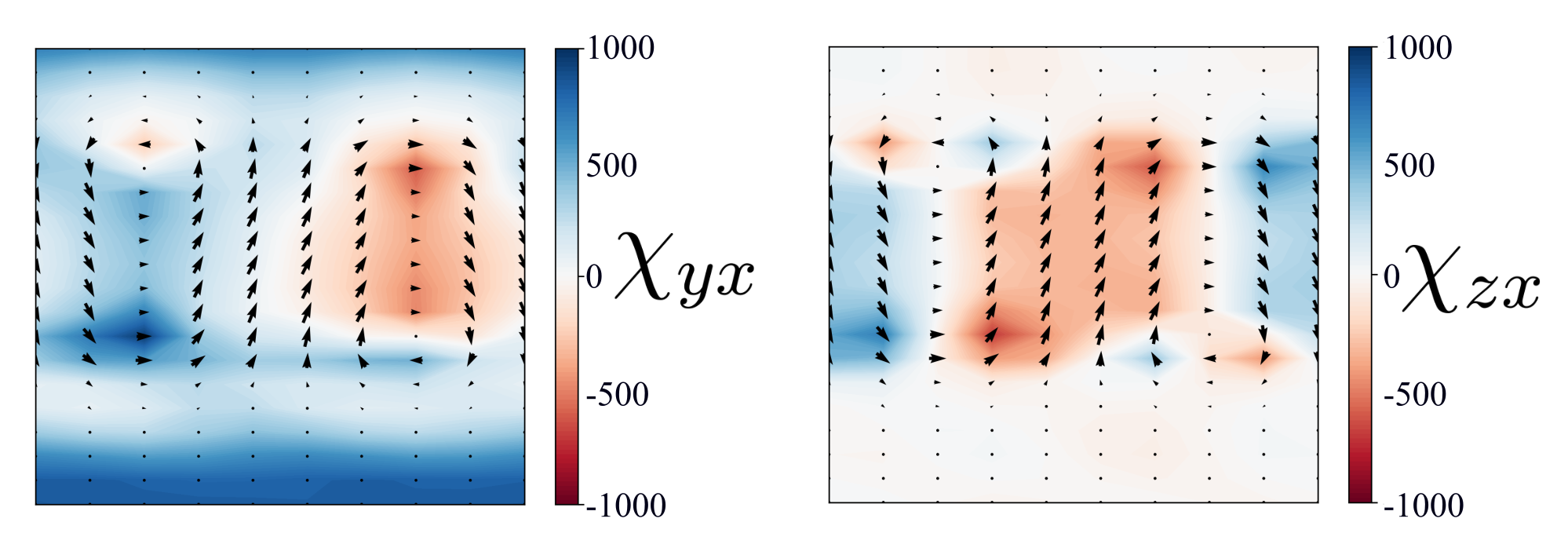}
      \caption{Transverse local distribution of electric susceptibility responses in the opposite buckling state as the one illustrated on Fig.~\ref{fig:Fig1} and Fig.~\ref{fig:Fig2}.}
      \label{fig:Fig3} 
\end{figure}
\begin{figure*}[tbhp]
     \centering
      \includegraphics[width=\textwidth]{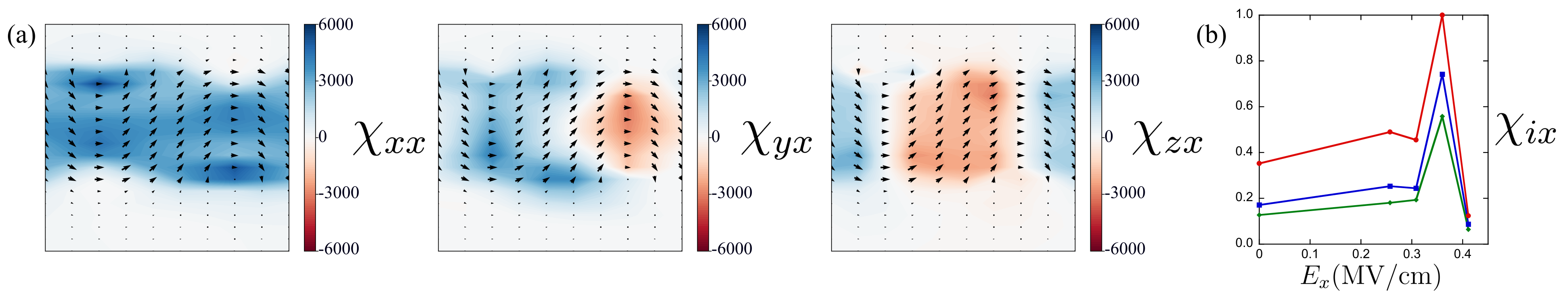}
      \caption{(a) Local distribution of electric susceptibility responses under an electric bias of $0.36$ MV/cm applied along the $x$-direction. (b) Dependence of the maximum susceptibility response as a function of the applied field $E_x$. Red, blue, and green curves represent the evolution of the $x$, $y$ and $z$ normalized susceptibility responses.}
      \label{fig:Fig4} 
\end{figure*}

Clearly, as a result of this observation, the local response is inherently dependent on the geometry of the sample and their collective dynamics~\cite{Li-21,Huaiyu-25}.
For instance, if we see the response on the SrTiO$_3$ layers that show a trivial non-topological behavior the dielectric tensor is homogeneous and typically lower in magnitude.
Moreover, if the buckling of the vortex cores is reversed, the associated local regions of positive and negative transverse susceptibility responses will likewise reverse their signs. This behavior is indeed confirmed in our simulations, as illustrated in Fig.~\ref{fig:Fig3}. 
Notably, the sense of the buckling is now reversed compared to Fig.~\ref{fig:Fig2}, as evidenced by the projection of the local dipoles onto the $(x,z)$ plane. As can be observed, the transverse susceptibility responses corresponding to the reversed buckling state exhibit a sign reversal while maintaining the same magnitude, as anticipated from the preceding discussions. This is to the best of our knowledge the first system where the electric susceptibility can be switched by an applied homogeneous electric field.

This observation confirms our hypothesis that the transverse susceptibility responses are intrinsically dependent on the geometry of the sample. Consequently, this suggests that such responses could exhibit significant changes during structural transitions within the material. For instance, it is well known that polar vortices can transit to polarization waves~\cite{Prosandeev-10} under electric fields or tensile strains along the $x$-direction. 
To explore the evolution of the transverse susceptibilities under such transformations, we performed finite difference calculations ($0.026$ MV/cm) about increasing background fields along the $x$-direction. We observed that under an electric bias of $0.36$ MV/cm the transition towards the polarization wave state occurs. At this point we observe an increase of the magnitude of the local transverse susceptibilities of 340\% that become comparable with the diagonal components as shown in Fig.~\ref{fig:Fig4}.

\begin{figure}[bhp]
     \centering
      \includegraphics[width=\columnwidth]{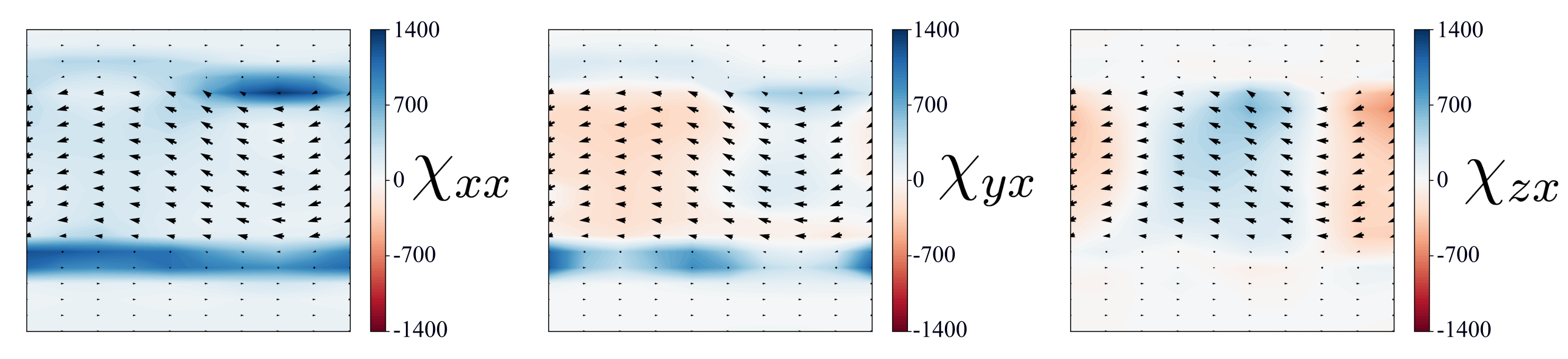}
      \caption{Local distribution of electric susceptibility responses computed through finite differences of the electric field when the system is subjected to a biaxial tensile epitaxial strain of $0.75\%$.}
      \label{fig:Fig5} 
\end{figure}

Interestingly, we find that a similar transition towards the polarization wave state can also be induced by increasing the tensile strain in the structure, even in the absence of an external electric background field. By gradually enhancing the in-plane lattice parameters, we observe a comparable transformation of the polarization texture. As shown in Fig.~\ref{fig:Fig5}, for a strain value of $\epsilon_{xx}=\epsilon_{yy}=0.75\%$ (corresponding to $a=b=3.930$ \AA) we find that the susceptibility responses $\chi_{xx}$, $\chi_{yx}$ and $\chi_{zx}$ become comparable. As it can be observed from the figure, the diagonal component $\chi_{xx}$ is mostly concentrated at the interface between the PbTiO$_3$ and SrTiO$_3$ layers whereas inside the PbTiO$_3$ layers the off-diagonal components $\chi_{yx}$ and $\chi_{zx}$ dominate. 

In summary, we predict the emergence of local transverse susceptibility responses within the polar vortex phase of PbTiO$_3$/SrTiO$_3$ superlattices. This novel dielectric behavior is characterized by polarization changes occurring perpendicular to the applied electric field, with magnitudes comparable to the regular (diagonal $\partial P_i/\partial E_i$) susceptibility terms, being approximately half of their value. 
Remarkably, the local transverse components of the susceptibility tensor are closely tied to the geometry of the polarization texture and their collective dynamics. Consequently, their magnitude varies significantly, by up to a factor of three, during the transition from the polar vortex phase to the polarization wave phase, driven by the applied electric bias. Furthermore, the sign of the local susceptibility can be reversed when the buckling between adjacent vortices is inverted, highlighting the tunable, deterministic and dynamic nature of these transverse responses under external homogeneous electric fields.
This is, to the best of our knowledge, the first reported system where highly tunable transverse local dielectric responses can be switched in a deterministic manner, providing a new avenue for designing devices with controllable and dynamic dielectric properties at the nanoscale. Moreover, this concept may be general, and similar behavior could potentially be observed in other topological textures where complex collective dynamic behavior occurs, as has been recently reported in polar skyrmions~\cite{Huaiyu-25}.
\acknowledgments
F.G.O. acknowledges financial support from MSCA-PF 101148906 funded by the European Union and the Fonds de la Recherche Scientifique (FNRS) through the grant FNRS-CR 1.B.227.25F and the European Union’s Horizon 2020 research and innovation program under Grant Agreement No. 964931 (TSAR). R.R. acknowledges support from the ONR-MURI ETHOS program under contract \textit{via }cooperative agreement W911NF-21-2-0162. JJ acknowledges financil support from Grant PID2022-139776NB-C63 funded by MICIU/AEI/ 10.13039/501100011033 and by ERDF/EU.
%
\onecolumngrid
\section*{End Matter}
\twocolumngrid
\emph{Computational Deails.-}
The second-principles simulations were performed using the same methodology presented in previous works~\cite{Wojdel-13,Zubko-16}, as implemented in the {\sc{scale-up}} package~\cite{Pablo-16}.
The interatomic potentials, and the approach to simulate the interface, are the same as in Ref.~\cite{Zubko-16}. We impose an epitaxial constraint assuming in-plane lattice constants of $a=b=3.911$ \AA,~forming an angle of $\gamma=90^\circ$. This corresponds to a small tensile epitaxial strain of +0.25\% with respect to the reference structure used in previous works (where $a = b = 3.901$ \AA) similar to the one used on Ref.~\cite{Behera-22,Behera-23}.

The supercell dimensions corresponds to 10$\times$2$\times$$\left[\right.$(PbTiO$_3$)$_{10}$/(SrTiO$_3$)$_{10}$$\left.\right]$. The lateral size along the $x$-direction is chosen as the optimal period for the particular supercell thickness following previous works~\cite{Gomez-23} whereas the $y$-directions is chosen to be $2$ unit cells. This size is sufficiently large to allow octahedral rotations in the SrTiO$_3$ layers, which can affect the dielectric response of the material. Additionally, limiting the structure to only $2$ unit cells along the $y$ direction helps speed up the calculations while maintaining the essential physics of the system since perfectly periodic stripe domains are expected at low temperatures as demonstrated in Ref.~\cite{Gomez-22}.

All the simulations performed in this study are conducted at $T=0$ K, we solved the models by running Monte Carlo simulations comprising 10,000 thermalization sweeps, followed by another 10,000 steps to compute local polarizations. Local polarizations are computed within a linear approximation of the product of the Born effective charge tensor times the atomic displacements from the reference structure positions divided by the volume of the 5-atoms unit cell centered on the A-sites of the perovskite.

To compute the local susceptibilities we followed a central-derivative finite differences approach and performed ten different realizations starting from the previously thermalized sample showing a well defined offset as the one depicted in Fig.~\ref{fig:Fig1} with an electric field of $\pm0.026$ MV/cm directed along $x,~y$ and $z$ directions (60 calculations in total). The atomic displacements of the 10 different realizations at each electric field where averaged and used to compute the local polarizations. Then, the local susceptibility components $\chi_{ij}(\mathbf{r})$ at a given cell site are computed as 
\begin{equation}
    \chi_{ij}(\mathbf{r})=\frac{1}{\varepsilon_0}\frac{P_i(\mathbf{r},+E_j)-P_i(\mathbf{r},-E_j)}{2\vert E_j\vert}.
\end{equation} 

\end{document}